\documentclass[]{aastex631}
\usepackage{xcolor}

\received{2024 December 21}
\revised{2025 February 13}
\accepted{2025 March 3}

\submitjournal{PSJ}

\shorttitle{Preliminary NEAT Results}
\shortauthors{Nugent et al.}
\graphicspath{{./}{}}

\begin{document}
\title{Reprocessing the NEAT Dataset: Preliminary Results}

\correspondingauthor{C. R. Nugent}
\email{cnugent@olin.edu}

\author[0000-0003-2504-7887]{C. R. Nugent}
\affiliation{Olin College of Engineering\\
1000 Olin Way \\
Needham, MA 02492, USA}

\author[0000-0001-9542-0953]{J. M. Bauer}
\affiliation{Department of Astronomy\\
University of Maryland\\
College Park, MD 20742, USA}


\author{O. Benitez}
\affiliation{Olin College of Engineering\\
1000 Olin Way \\
Needham, MA 02492, USA}

\author{M. Blain}
\affiliation{Department of Earth and Space Sciences\\
College of the Canyons\\
Santa Clarita, CA 91355, USA}

\author{N. D'Souza}
\affiliation{Olin College of Engineering\\
1000 Olin Way \\
Needham, MA 02492, USA}

\author{S. Garimella}
\affiliation{Olin College of Engineering\\
1000 Olin Way \\
Needham, MA 02492, USA}

\author[0000-0003-3369-7097]{M. Goldwater}
\affiliation{MIT-WHOI Joint Program in Applied Ocean Science \& Engineering, Cambridge and Woods Hole, MA, USA}
\affiliation{Department of Applied Ocean Physics \& Engineering\\
Woods Hole Oceanographic Institution\\
266 Woods Hole Road\\
MS \#55\\
Woods Hole, MA 02543-1050, USA}

\author{Y. Kim}
\affiliation{Department of Astronomy\\
University of Maryland\\
College Park, MD 20742, USA}

\author[0009-0001-4346-381X]{H. C. G. Larsen}
\affiliation{Aalborg University\\
Fibigerstr\ae de 16\\
9220 Aalborg, Denmark}

\author[0000-0003-2534-673X]{T. Linder}
\affiliation{Planetary Science Institute,
Tucson, AZ 85719 USA}

\author{K. Mackowiak}
\affiliation{Olin College of Engineering\\
1000 Olin Way \\
Needham, MA 02492, USA}

\author{Z. McGinnis}
\affiliation{Olin College of Engineering\\
1000 Olin Way \\
Needham, MA 02492, USA}

\author{E. Pan}
\affiliation{Olin College of Engineering\\
1000 Olin Way \\
Needham, MA 02492, USA}

\author[0009-0002-7177-0638]{C. C. Pedersen}
\affiliation{Aalborg University\\
Fibigerstr\ae de 16\\
9220 Aalborg, Denmark}

\author{P. Sadhwani}
\affiliation{Olin College of Engineering\\
1000 Olin Way \\
Needham, MA 02492, USA}

\author{F. Spoto}
\affiliation{Minor Planet Center\\
Smithsonian Astrophysical Observatory\\
60 Garden Street\\
Cambridge, MA 02138, USA}

\author[0000-0001-6541-8887]{N. J. Tan}
\affiliation{ School of Physical and Chemical Sciences — Te Kura Mat\=u\\
University of Canterbury\\
Private Bag 4800\\
Christchurch 8140, New Zealand}

\author{P. Vere\v{s}}
\affiliation{Minor Planet Center\\
Smithsonian Astrophysical Observatory\\
60 Garden Street\\
Cambridge, MA 02138, USA}

\author{C. Xue}
\affiliation{Olin College of Engineering\\
1000 Olin Way \\
Needham, MA 02492, USA}

\begin{abstract}

We have created a new image analysis pipeline to reprocess images taken by the Near Earth Asteroid Tracking survey and have applied it to ten nights of observations. This work is the first large-scale reprocessing of images from an asteroid discovery survey in which thousands of archived images are re-calibrated, searched for minor planets, and resulting observations are reported to the Minor Planet Center. We describe the software used to extract, calibrate, and clean sources from the images, including specific techniques that accommodate the unique features of these archival images. This pipeline is able to find fainter asteroids than the original pipeline. 

\end{abstract}


\section{Introduction} \label{sec:intro}

Asteroid discovery surveys image large swaths of the sky, discovering hundreds of near-Earth objects (NEOs) each month \citep[e.g.][]{2015Jedicke}\footnote{Current monthly discovery rates can be found on the Minor Planet Center home page,  \url{https://minorplanetcenter.net/}.}. Comparison of discovery statistics with population models \citep[e.g.][]{2018Granvik, 2024Nesvorny} indicates that together, surveys have made significant progress toward discovering the NEO population larger than 140 m in diameter.

However, there is a portion of the NEO population that these surveys cannot detect on any given night: objects with locations relative to survey telescopes and the Sun that place them in the daytime sky. Over months or years, many NEOs drift out of this unobservable group and into the night sky, but some linger for decades. For example, consider the edge case 2002 AA29. Observations of this NEO were reported to the Minor Planet Center (MPC) from 2002 until 2004, after which it became unobservable. Because it has an orbital period of 0.99 years, it will remain unobservable from Earth until 2093.

To extend observed arcs of NEOs that linger in the daytime sky, researchers can seek precovery observations in archival data \citep[e.g.][]{2001Boattini, 2012Gwyn, 2013Vaduvescu, 2023Perlbarg, 2023Saifollahi}. Precovery observations can secure orbits, which is particularly valuable for objects that make close approaches to planets, including Earth.  However, archival data can contain artifacts that are uncommon in modern data, requiring special care and ruling out the use of some software tools. Additionally, undocumented issues related to the images, such as telescope movement or timing errors, can result in erroneous results. Extracting quality data from archival images therefore requires significant time, and in some cases, new software development.

We describe a new image processing pipeline applied to the archived Near Earth Asteroid Tracking (NEAT) survey, which operated from 1995 to 2007 and was led by Eleanor ``Glo'' Helin \citep{1990Helin, 1997Helin}. NEAT was a project of NASA’s Jet Propulsion Laboratory (JPL) and the Air Force Space Command. It employed several 1-meter class telescopes with CCD cameras and large fields of view. Together with Spacewatch \citep{1982Gehrels, 1986Gehrels, 1991Rabinowitz}, NEAT was one of the first asteroid surveys to use digital imaging technology and search for asteroids using computers. It began surveying from Haleakala, Maui, HI, with the Ground-based Electro-Optical Deep Space Surveillance (GEODSS) camera. The telescope was remotely accessed from Pasadena, California. Data were sent from the telescope to the computing resources at JPL using a modem and a 1-800 telephone number with an effective transmission rate of 1.5 kbps. In 2001, the survey started operating at Palomar Observatory Oschin Schmidt Photographic Equatorial Telescope in Southern California using the TriCam camera.  During its lifetime, NEAT discovered 41,227 minor planets and submitted over a million observations to the MPC.

The GEODSS camera was initially paired with the U.S. Air Force one meter telescope; in 2000 it was moved to the Maui Space Surveillance Site 1.2 m telescope\footnote{Planetary Data System; \url{https://pds.nasa.gov/ds-view/pds/viewContext.jsp?identifier=urn\%3Anasa\%3Apds\%3Acontext\%3Ainstrument\%3Aamos.neat_maui_camera&version=1.1}}. GEODSS had $4096 \times 4096$ pixels, with a 1.43 arcsecond/pixel scale on the first telescope and a 1.36 arcsecond/pixel scale on the second telescope. The CCD was read out by four separate amplifiers. \citet{1999Pravdo} reports a limiting magnitude of $19.1 \pm 0.1$. Images were taken without filters. Inspection of a GEODSS image taken in 1995 shows a FWHM of 2.6 pixels.

As its name implies, the TriCam camera consisted of three distinct $4096 \times 4096$ pixel CCD sets, or cameras.\footnote{Planetary Data System; \url{https://pds.nasa.gov/ds-view/pds/viewContext.jsp?identifier=urn\%3Anasa\%3Apds\%3Acontext\%3Ainstrument\%3Apalomar.oschin\_schmidt\_1m2.tricam&version=1.0}} 
The cameras are arrayed vertically and distinguished via `a', `b' and `c' labels. The survey took three exposures per field, the same camera was used for each exposure in a field. Each camera was read out by four amplifiers. The pixel scale for each camera is 1.01 arcseconds/pixel, therefore, each camera covers a bit more than a square degree. Images were taken without filters. The FWHM of sources is 3.7 pixels.

A typical NEAT survey cadence is described in \citet{1999Pravdo}. However the archived data \citep{NEATSBN} do not always match that described cadence. Here we provide details on the NEAT observations reprocessed in this work. These numbers may have changed over the course of the NEAT TriCam survey and cannot be assumed to apply to all NEAT TriCam observations. Between 120 and 462 fields were observed each night. Observing cadence information is available in the logs folder associated with each archived NEAT night in the Planetary Data System archive \citep{NEATSBN}. According to those log files, observations generally had 60 s exposures with either 900 or 1800 s between exposures in the same field. However, some nights had 150 s exposures with 5400 s between exposures. Other nights consisted solely of 20 s exposures, either with 900 or 1800 s between exposures.

The exposure time information in the PDS log files correlates with the exposure times recorded in the individual {\tt\string FITS} file headers. However, the time between exposures as determined by the differences between the {\tt\string  DATE\-OBS} field of the {\tt\string FITS} files and the times recorded in the PDS log files can be different by tens of minutes. We infer that the round number times recorded in the PDS log files were the planned cadence, and sometimes complicating real-world factors resulted in the executed cadence being different.

In the decades since NEAT operated, there have been improvements in computing hardware and software. We have created a new data analysis pipeline to reprocess the NEAT dataset to find new observations of minor planets in the images, extending the observed arcs. The original NEAT data processing pipeline was not archived and therefore could not be used, although parts are described in the appendix of \citet{1999Pravdo}. Although the new pipeline described in this paper was created to reprocess NEAT data, it has been structured so that it may be applied to other datasets.

\section{Methods} \label{sec:methods}

In this section, we describe the image calibration, source extraction, source screening, and source linking that comprises the backbone of the reprocessing effort. As nights of data have been reprocessed, there have been modifications to the parameters and minor improvements. This section describes the parameters used in the most recent reprocessing.

The pipeline is written in Python 3.11. Compressed images, in {\tt\string fits.fz} format, were downloaded from the Planetary Data System \citep{NEATSBN} using {\tt\string wget}. The pipeline processes one night at a time. Each full night of observations was processed sequentially on a single core, while multiple nights were processed concurrently using multiple cores. The code is structured with a processing class and a configuration dictionary. The dictionary includes options for displaying diagnostic images, parameters related to observations (such as the observatory code), and other variables discussed below. Processing was completed using the Massachusetts Green High Performance Computing Center operated by Boston University. 

\subsection{Initial Calibration}
\begin{figure}[ht!]
\plottwo{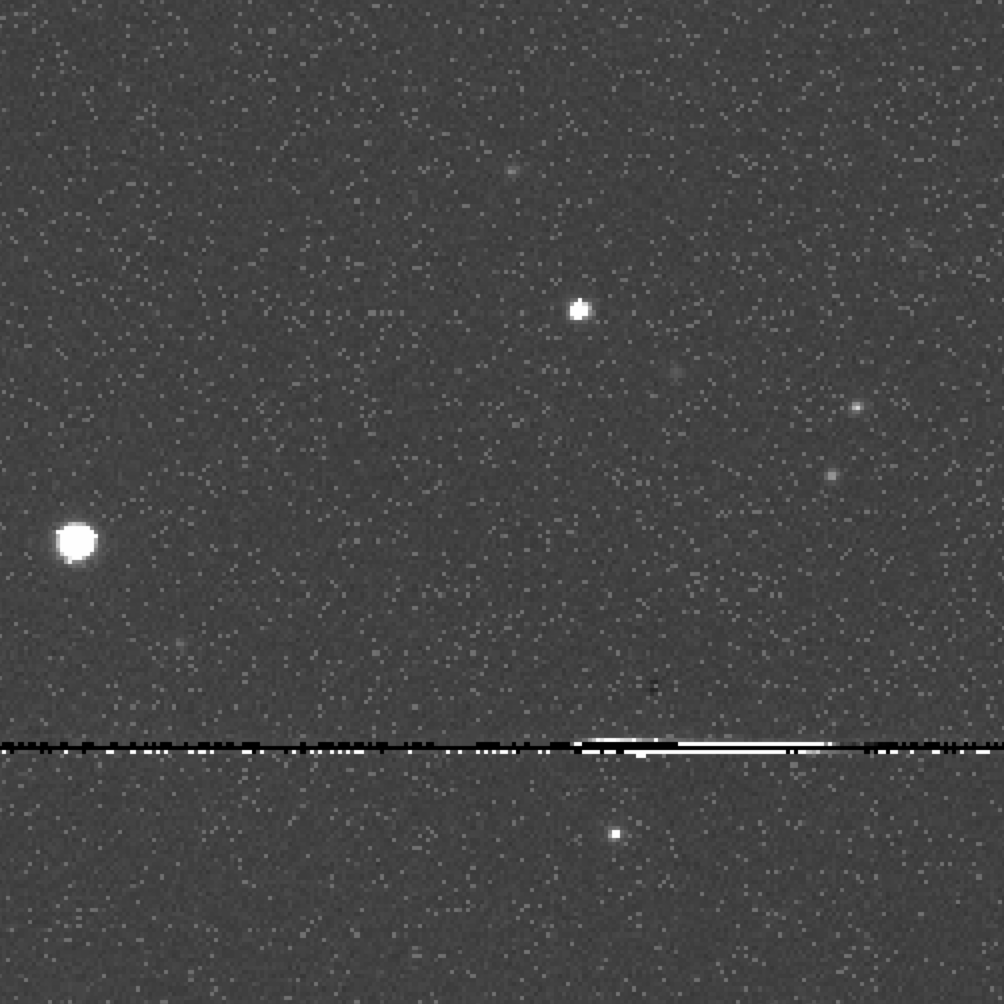}{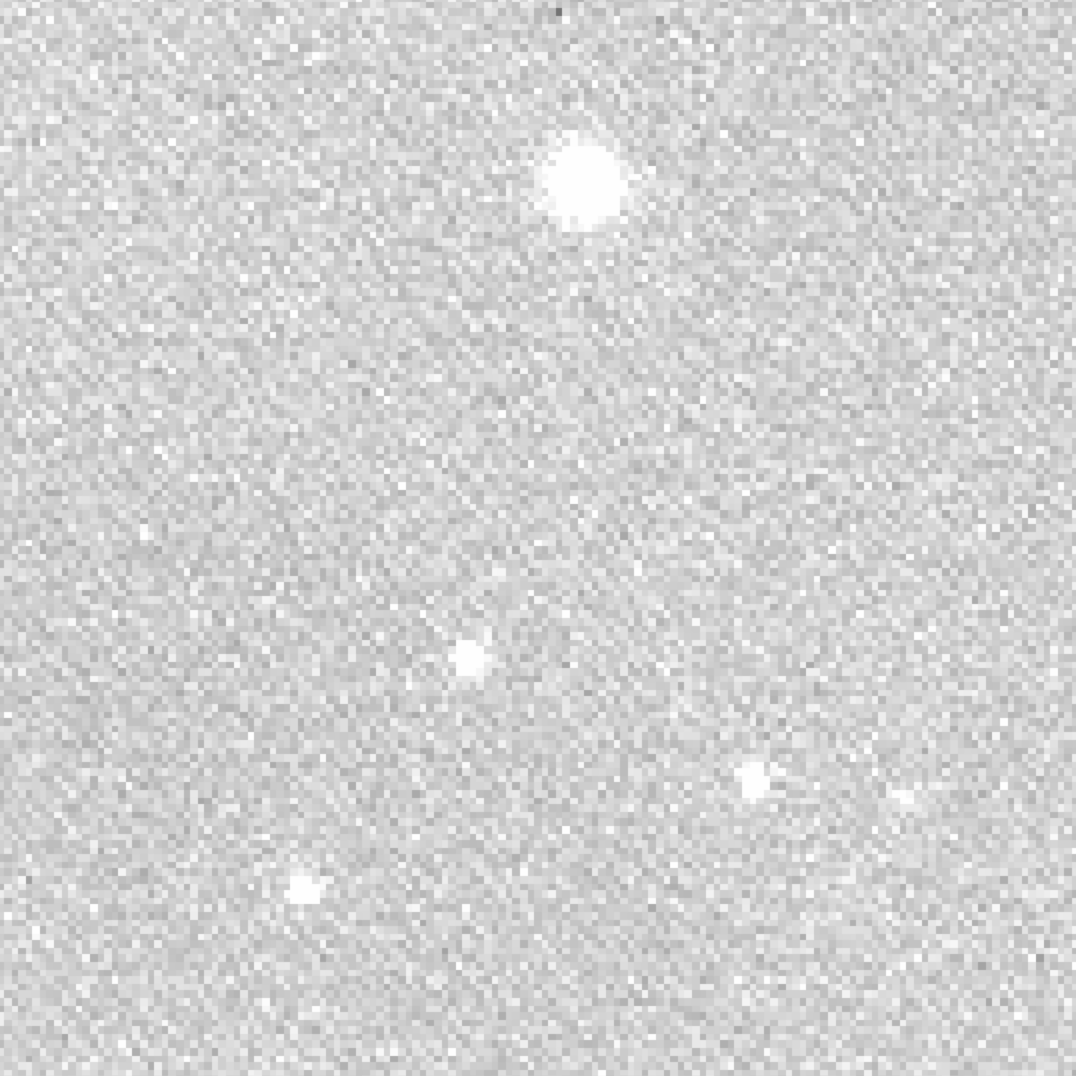}
\caption{Two regions from a TriCam `a' camera image taken on Nov 20 2001. These regions are from two different CCD regions on the same camera, and show characteristic noise and artifacts. The NEAT TriCam images are morphologically different than modern CCD images, with more noise and artifacts. \label{fig:orig_images}}
\end{figure}

Each image was treated as follows. Images were unpacked from {\tt\string fits.fz} format using the Astropy package \citep{2022Astropy, 2018Astropy, 2013Astropy}, and the image and header values were read in using  {\tt\string astropy.io fits.open}. In some cases, images were found to contain negative or infinite pixel values. To avoid issues with the source detection algorithm, these pixels were identified and reset to zero. As TriCam detectors were early examples of CCD technology, they contain artifacts and noise that are less common in modern telescopic images. Two different sections of unprocessed images are shown in Figure \ref{fig:orig_images}.

As each NEAT TriCam image was read out by four separate amplifiers, the image is split into four quadrants for background subtraction, source identification, and calibration. 

\begin{figure}[ht!]
\plottwo{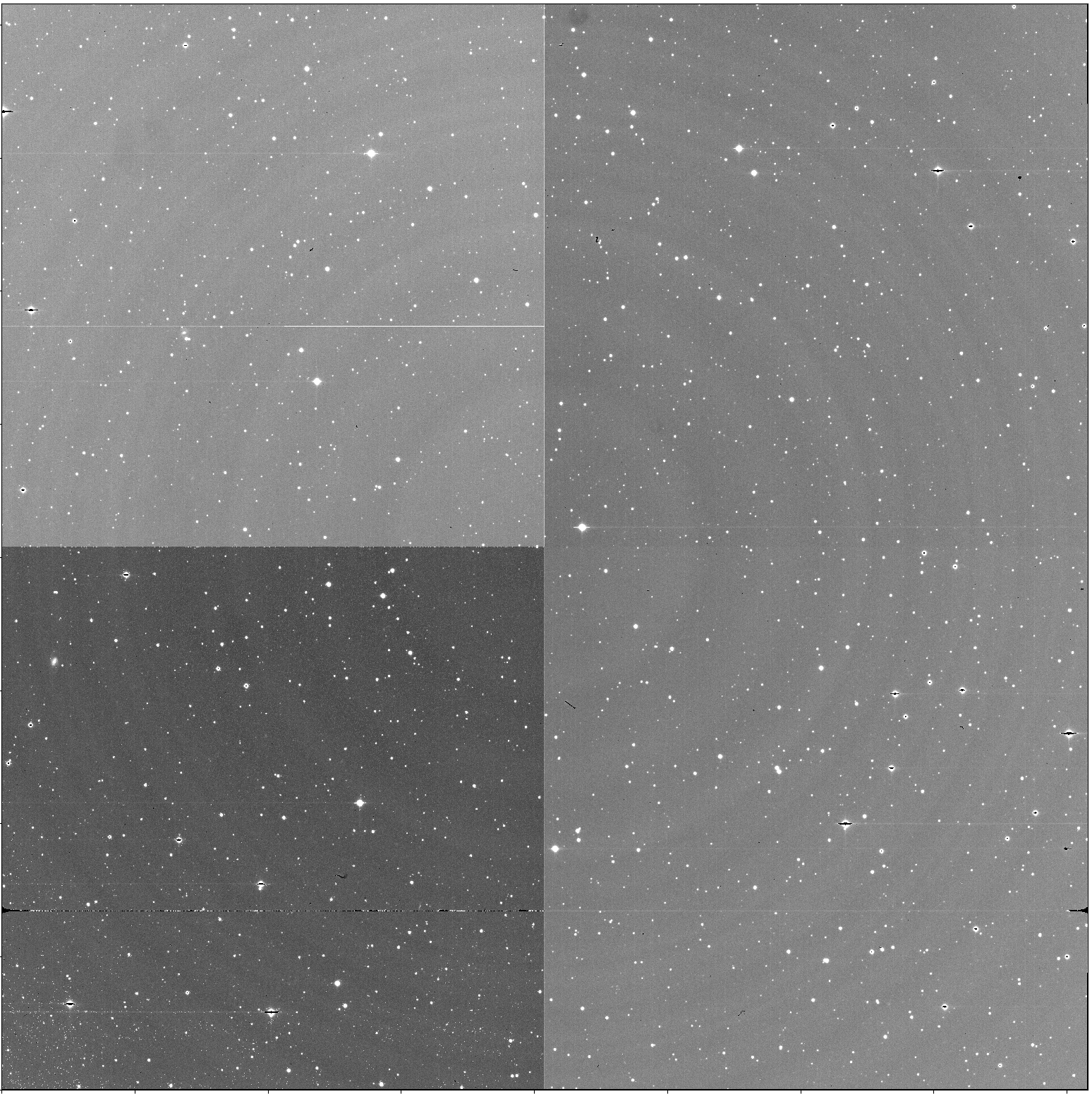}{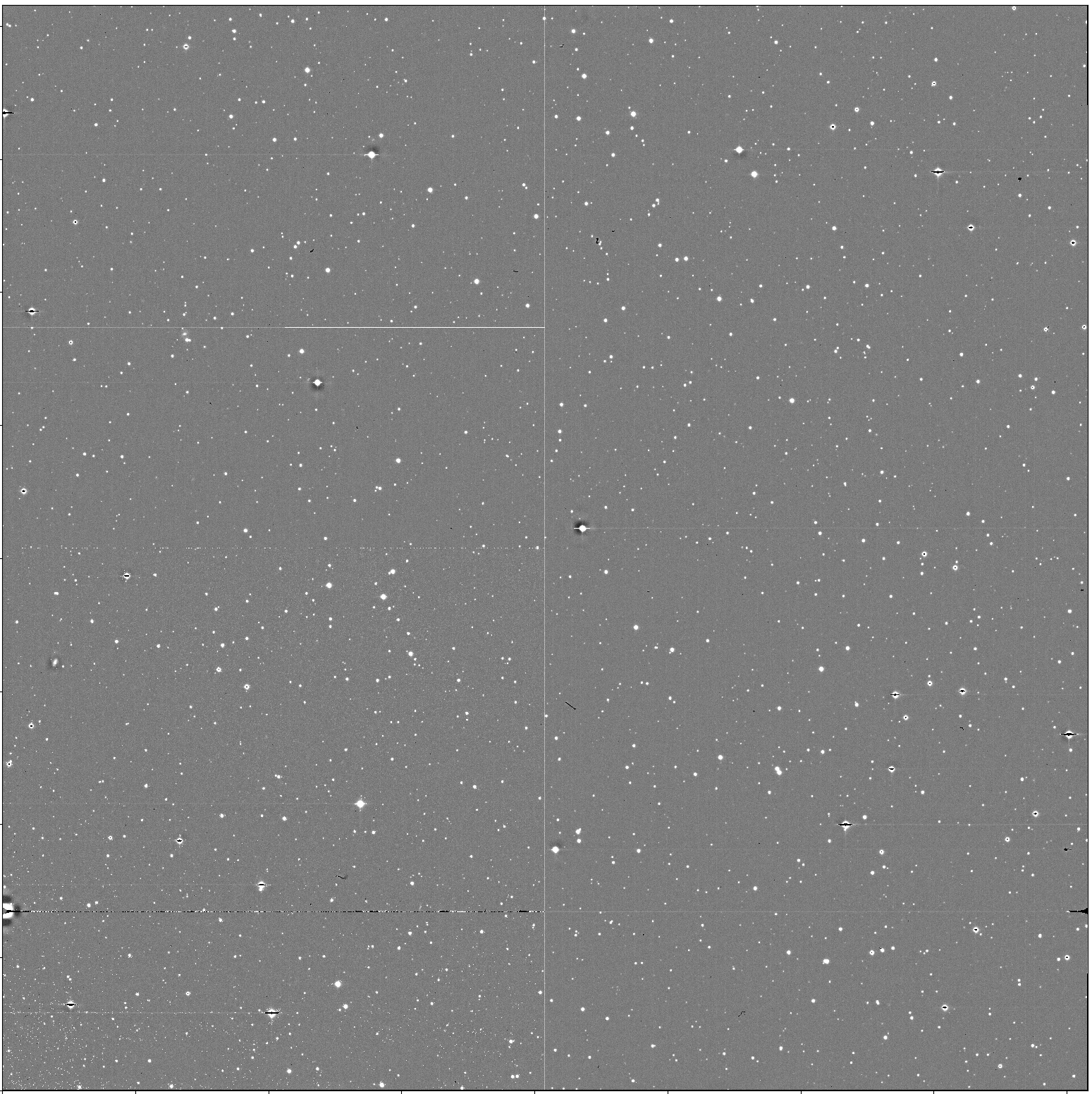}
\caption{A NEAT TriCam image frame before (left) and after (right) background subtraction. Algorithmic background subtraction was chosen as it produced uniform results over a range of frames and allowed for the detection of faint sources. \label{fig:bkgsub}}
\end{figure}

The background is subtracted using the  {\tt\string Background2D} function of {\tt\string photutils}, a photometry package that is affiliated with Astropy \citep{2023Bradley}. Results are shown in Figure \ref{fig:bkgsub}.  This method was selected after testing several alternative functions available in various libraries. It proved robust enough to handle the particular noise of the TriCam CCDs. Image calibration using the archived NEAT dark and flat exposures was tested, but was not implemented as it produced noisier results than algorithmic background subtraction. Masking of dead pixels by value was tested, but not implemented in the final pipeline as it did not significantly improve source detection. 

\begin{figure}[ht!]
\plotone{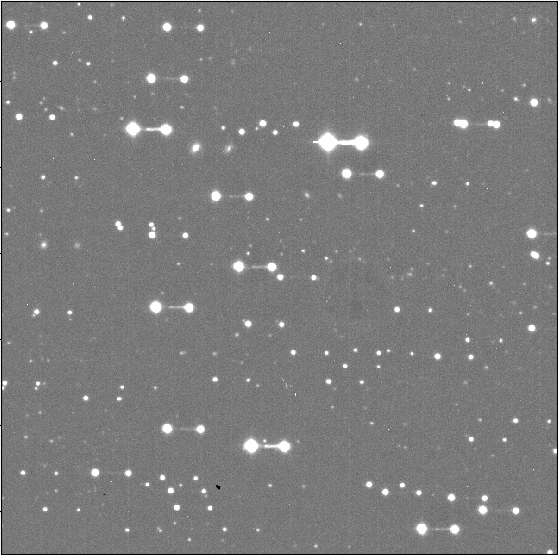}
\caption{An example of a frame that experienced telescope movement during the exposure. These frames were identified by the pipeline and ignored. \label{fig:double}}
\end{figure}

Some archived NEAT TriCam frames suffered from horizontal movement during the exposure (Figure \ref{fig:double}). This was discovered during this work, and we were not able to find previous documentation of the issue. Before processing, frames are checked for this movement. The {\tt\string DAOStarFinder} class is used to identify sources; this is a separate process than the main source extraction routine described in Section \ref{subsec:sourceid}. The brightest $30\%$ of sources are isolated. For each bright source, we consider a circular area centered 30 pixels above of each of these bright sources with a radius of 25 pixels, and count the number of sources inside. The total number of sources in the circle is counted, not just the bright sources. This is compared to similar circles to the left and right of the bright sources. If either the left or the right circles have 1.25 times the number of sources than the circle above, the frame is considered a double exposure, and the pipeline moves on to the next frame. 

\subsection{Source Identification}\label{subsec:sourceid}

Sources were identified using the {\tt\string DAOStarFinder} class, part of the {\tt\string photutils} package; this is based on {\tt\string DAOFIND} \citep{1987Stetson}. This algorithm was chosen after testing of alternative source detection routines, including SExtractor \citep{1996Bertin}; {\tt\string DAOStarFinder} was shown to identify faint sources with high accuracy while not producing a large excess of false sources due to the CCD readout boundaries or CCD saturation features adjacent to bright stars. The parameters that were used for {\tt\string DAOStarFinder} included sharpness and roundness ranges broader than the usual default. Sharpness ranged from $0 - 2.5$ and roundness from $-1.5$ to $1.5$.

Sources are first extracted using a signal-to-noise (SNR) ratio threshold of 4. If the number of sources in the quadrant exceeds 10,000, then the threshold is increased by 1, and the sources are extracted again. This process repeats until the number of extracted sources is less than 10,000. This process is needed because sensitivity varies between CCD regions, and allows for adjustment in extraction due to time-dependent variables such as changing cloud cover. When a single SNR of 4 was used for all data, the number of sources extracted was too high, confounding the detection-linking software later on the process. A fixed SNR value of 5 or 6 resulted in some real, faint sources in sensitive chip regions not being detected. In practice, the SNR usually varied between 4 and 5.

A circular aperture was used to calculate aperture photometry via the {\tt\string photutils} package. The resulting list of sources was converted to a {\tt\string pandas DataFrame};  {\tt\string pandas} \citep{2010mckinney,pandas} is used throughout the remainder of the pipeline to manipulate the resulting data. Although other data formats, such as Apache Parquet, are faster, {\tt\string pandas} was sufficient for this project's needs.

\subsection{Astrometric and Photometric Calibration}

The archived  {\tt\string fits.fz} images include telescope pointings as part of the World Coordinate System (WCS) header information. However, the accuracy of that WCS is not sufficient for minor planet astrometry. We therefore needed to perform further astrometric calibration employing a modern reference star catalog, Gaia DR3.

While the original NEAT pipeline for the GEODSS telescope was described in \citet{1999Pravdo}, we have been unable to locate documentation for the TriCam image pipeline. As previously noted, the TriCam camera was used with the Oschin Schmidt Photographic Equatorial Telescope at Palomar Observatory. Schmidt telescopes have significant on-plate distortion.
 We inferred from star locations that perhaps there are physical offsets between the four amplifier-readout quadrants. On each of the `a', `b', and `c' cameras, one quadrant sometimes appears to be spatially located farther from the center of the image than the others, with more telescope-induced distortion on the corner farthest from the center. Therefore each image quadrant was individually calibrated.

The following astrometric and photometric calibration process was selected after trying several combinations of routines. This process was found to be robust to the wide variety of exposures and artifacts in the NEAT TriCam images. 

We query the  {\tt\string Astrometry.net} calibration service \citep{2010Lang} via the {\tt\string solve\_from\_source\_list} function of the {\tt\string Astroquery} library \citep{2019Ginsburg}. This query supplied an initial guess of the midpoint of the quadrant (taken from the telescope pointings recorded in the header) and the 300 brightest sources found by  {\tt\string DAOStarFinder}. The query, when successful, returned a WCS header corresponding with with improved astrometry. The WCS values used by the pipeline were updated with these new results.

In some cases, {\tt\string Astrometry.net} could not successfully solve a frame quadrant. Sometimes this was due to the pipeline overloading the servers with queries, therefore the code would first try waiting ten minutes before resubmitting the query. If the query failed a second time, the pipeline recorded the failure and moved on to another frame. Frames that failed  {\tt\string Astrometry.net} were uniformly found to lack visually apparent stars.

We queried the Gaia DR3 release for the stars in the region. Frame corners were determined from the WCS returned by {\tt\string Astrometry.net}. All the stars brighter than 20.7 mag in the frame region were retrieved via {\tt\string Astroquery}, which queries the Gaia Table Access Protocol (TAP+)  for the Gaia DR3 release. Gaia stars' positions are given at the reference epoch 2016.0 \citep{2023Gaia}. That reference epoch is precessed to J2000 in the ICRS reference frame using the same TAP+ interface during this query.  After the query, a proper motion correction was applied to the stars, so that the stars are registered in the correct locations at the time of the image exposure, which is taken from the  {\tt\string fits.fz} header information. 

The NEAT images present some challenges when it comes to astrometric calibration. Due to the Schmidt telescope, the corners of the images display significant distortion, resulting in up to 10''-15'' displacements between image sources and their corresponding reference stars. This distortion cannot be modeled well by a TAN WCS projection. \citet{1999Pravdo} warn against using a distortion map to correct for this distortion on NEAT GEODSS images, writing, ``The important aberration is distortion whose affect is to cause a cubic deviation from linearity with radius of object positions. While the magnitude of this term is well-measured, the phase depends upon the center of the field, which changes depending on slight differences in the mounting of the camera on the telescope (of order 100 $\mu m$).'' We interpret this sentence as advising against creating a generalized mapping of the distortion of telescope by computing the distortion for several images and averaging the result to create a distortion template that is then applied to all images as a calibration starting point. This interpretation is consistent with the results of a test we performed which created and applied such a distortion template as a initial correction to the images. This routine did not effectively improve the astrometry and was discarded.

Stars were associated with detected sources based on sky position. Throughout the pipeline, it was frequently necessary to spatially match two groups of data. To do this, we employed a Ball Tree data structure, implemented using the {\tt\string scikit-learn} module \citep{2011Pedregosa}. This data structure was chosen because it provided advantages in terms of speed while preserving accuracy. The haversine distance metric was used for fast association of large numbers of sources, though the resulting distance measurement of the haversine approximation can be inaccurate at small distances (roughly less than 10 arcseconds). When accuracy was needed in associations, but only a small number (less than 500) of associations needed to be made, a custom distance function that used Astropy's {\tt\string separation} function was used instead to assemble the Ball Tree. When both speed (involving greater than 500 associations) and accuracy (less than 10 arcseconds) was needed, we used a hybrid approach, where the haversine metric provided a first guess that was then checked using the {\tt\string separation} function. 

To correct for the distortion across each CCD quadrant, we first used a subset of reference stars between 13 and 15 mag. This range was selected because it reliably produced enough stars to cover the extent of each quadrant while being sparse enough that most stars would be associated with the correct sources. The brightest $15\%$ of sources were used to match with the subset of reference stars. Sources and stars were associated if they were within 15 arcseconds of each other. This was an experimentally determined number.  This value, although large, was needed to successfully match stars with sources at the edges of the detectors. However, this large search radius results in roughly $10\%$ of the star-source associations being spurious. Across a quadrant, correct star-source matches are similar to their neighbors in terms of direction and magnitude of displacement. The {\tt\string sklearn} function {\tt\string LocalOutlierFactor} employs that similarity between neighboring points to identify outliers, which were removed.  A degree 3 polynomial was fit to the remaining source-star matches, and the resulting fit was applied to all sources.

This process was then repeated using a subset of reference stars between 15 and 17.5 mag. This subset of stars more thoroughly covers each quadrant, and after the previous correction a smaller association radius of 5 arcseconds could be used. Again, the  {\tt\string LocalOutlierFactor} function was used to identify and remove outliers, and a degree 3 polynomial was fit to the source-star matches, and the resulting fit was applied to all sources. This was the final step in the astrometric calibration process.

We measured the quality of our astrometric calibration via a separate validation routine, using a different subset of stars, after the calibration was completed. We selected stars between 17.5 and 17.8 mag and associated them with sources using an association radius of 10 arcseconds.  The resulting star-source pairs were not cleaned with {\tt\string LocalOutlierFactor}. The offset between each pair in RA and Dec was recorded, and the standard deviations of offsets in RA and Dec were calculated. These standard deviations were used for the reported astrometric uncertainty. In rare cases, the calculated offsets were on order 0.1 arcseconds, due to the distribution of stars in the validation subset not covering the full extent of the frame. We therefore  impose a minimum uncertainty of 0.5 arcseconds, which captures the accuracy generally achievable by the pipeline. 

Gaia stars between 15 and 17 mag are used for photometric calibration. Gaia g-band measurements were used, namely the {\tt\string phot\_g\_mean\_mag} column. The headers for the NEAT TriCam {\tt\string fits.fz} images do not contain any zero point information. Stars were once again associated with sources, a threshold of three arcseconds was required to establish an association. Associations were used to calculate a zero point, which was then applied to all detected sources.

\subsection{Data cleaning and classification}

Stationary sources were removed via their positions in the sky. All detected sources were compared to all Gaia stars predicted to be in the frame. If a source was within 3 arcseconds of a star, it was removed. This removal radius was used because bright sources in the NEAT images span 6 or more pixels, and {\tt\string DAOPHOT} can sometimes produce multiple detections inside very bright sources. 

Sources are then screened via a supervised machine learning model. This classifier seeks to distinguish between minor planets and other image features that were detected by the pipeline. It operates on 17x17 pixel (corresponding with 17.17'' x 17.17'') thumbnails centered at the source location. Thumbnail pixel values were scaled between 0 and 1. 

A training set for the model was created as follows. Thumbnails were created around the positions of all known solar system objects visible in a randomly selected single night. The positions of the known solar system objects were retrieved using queries to the  {\tt\string SkyBoT} service \citep{2006Berthier} via  {\tt\string astroquery} \citep{2019Ginsburg}. These thumbnails were screened by visual inspection to ensure that a source was visible in the thumbnail. 

This training set did not explicitly include trailed objects. The morphologies of the known objects were diverse due to the variations across chips used in the separate cameras and the distortion of the telescope; point sources can appear diffuse and non-streaked objects can appear elongated. Therefore it is possible that slightly trailed detections would be recognized by the classifier.

\begin{figure}[ht!]
\plottwo{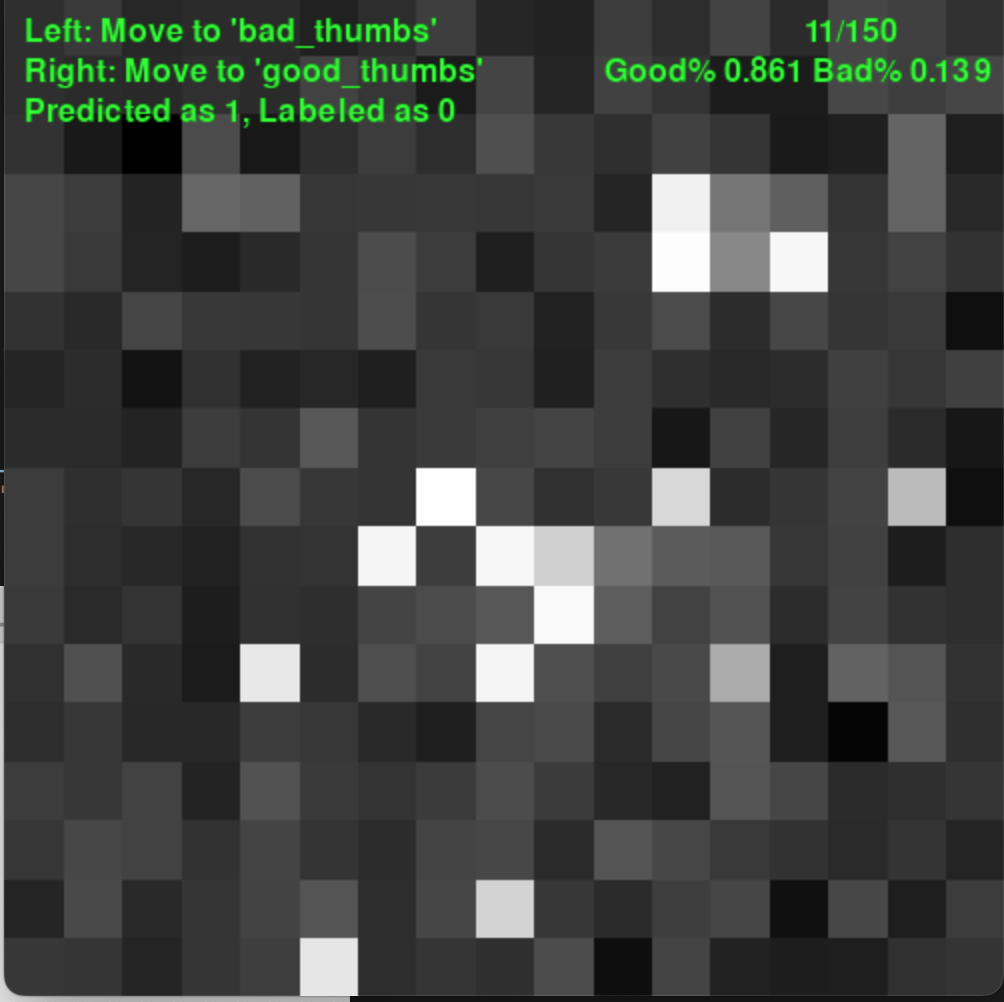}{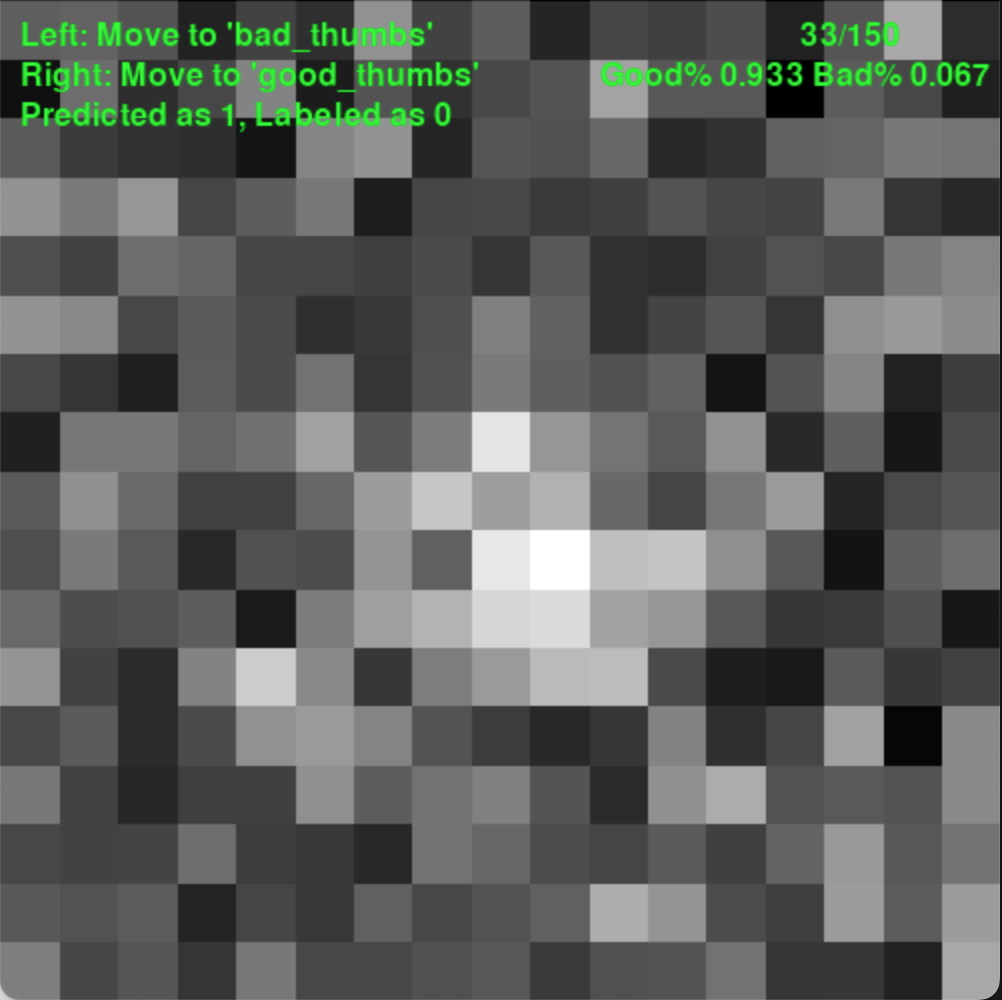}
\caption{Example output of helper software used to create and evaluate the machine learning model. A false positive (left) and an example of a misclassified positive source (right) are shown. Users can use the left or right arrow keys to reclassify the images, if desired. 
\label{fig:ml_helper}}
\end{figure}

 A training set of non-solar system objects was assembled by visual inspection of candidate thumbnails that were not associated with known sources. The total training set size was $\sim5000$ images. To assist in thumbnail sorting and training set quality control, helper software was written so that a user could sort thumbnails using keyboard arrow keys. Software to check false positives and false negatives, which also allowed for corrections of classification mistakes, was also created (Figure \ref{fig:ml_helper}).

\begin{figure}[ht!]
\plotone{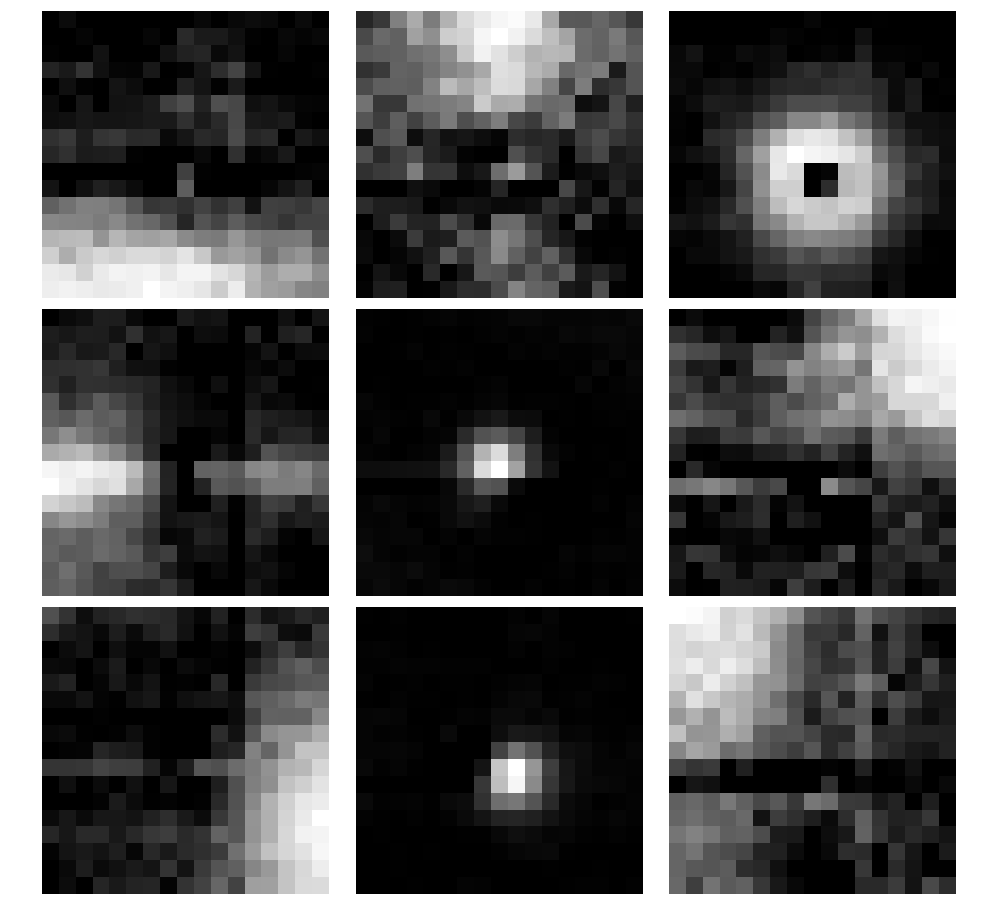}
\caption{The nine features identified by the machine learning model. This set of images represents how the thumbnails were divided into regions by the algorithm to allow for classification. To be classified as a positive source, an image needed to have bright pixels corresponding to the bright regions in the center middle and bottom middle squares, for example.  This represents the model used; similar models with slightly more or fewer features would have also been successful for the classification used in this work. 
\label{fig:ml}}
\end{figure}

Several machine learning models were tested, including neural networks. A model combining non-negative matrix factorization and a support vector machine methods was selected. The implementation employed the {\tt\string sklearn} module \citep{2011Pedregosa}. The features used by this model can be seen in Figure \ref{fig:ml}.  Testing on a validation set indicated $96\%$ recall of the model. As the model was applied to each night of data, false positives and negatives were identified, the training set was updated, and the model retrained. This process is known as hard negative and hard positive mining.

In the pipeline, the model assigns a probability that the thumbnail contains a solar system object. Sources are then screened on that probability before linking. 

The  {\tt\string DataFrame} of sources is exported as a comma-separated values (csv) file. Metrics associated with the processing, including number of sources found, the magnitude of the faintest and brightest source, average brightness of all sources in the frame, and processing time were recorded in a separate file. 

\subsection{Tracklet assembly and MPC submission}

These source {\tt\string csv} files are then grouped into triplets; the NEAT cadence took three images of the same part of the sky for each night. The Small Bodies Node archive\footnote{\url{https://pdssbn.astro.umd.edu/}} contains a {\tt\string tab} file for each night that records which images comprise a triplet. These source triplets are input into the linking software {\tt\string FindPOTATOs}  \citep{2025Nugent} to produce candidate tracklets, which are groups of three detections of a single object. Tracklet formation is a necessary requirement for data to be accepted by the MPC\footnote{\url{https://minorplanetcenter.net/}}, the clearinghouse and international archive of minor planet observations. 

{\tt\string FindPOTATOs}  rejected sources that were within 1.5 arcseconds of another source in the image triplet as stationary. A maximum speed of 0.1 arcseconds/second was imposed on the tracklets, higher than the threshold of 0.05 arcseconds/second seen in MPC-reported  tracklets from NEAT in 2001 and 2002. In some cases, NEAT observed regions of the main belt with a slower cadence; we therefore also imposed a limit of 60 arcseconds as the maximum distance an object can travel between frames. This is necessary to prevent a failure mode in the linking code, which happens when all sources can be linked to all other sources in the frame. The assembled tracklets were rejected if the total magnitude variance was greater than 1.5, and if the angle described by the three tracklet points was less than $150^\circ$.

Sometimes two or more tracklets rely on a single detection. The results are cleaned of these duplicates using {\tt\string check\_repeats.py} of {\tt\string FindPOTATOs}. This script assists the user in choosing which, if any, tracklets containing a duplicate detection to keep. Tracklets are then divided into re-measurements of previously submitted NEAT tracklets and candidate new tracklets. Both kinds of tracklets are reported to the MPC in ADES format\footnote{\url{https://github.com/IAU-ADES/ADES-Master/}}, which allows for inclusion of uncertainties in position and brightness measurements. The file name of the original image, as archived in the SBN Small Bodies Node, is given in the “Notes” field of the ADES submission for each observation. 

New candidate tracklets produced by {\tt\string FindPOTATOs}  are visually inspected for accuracy. Although point sources can register on the NEAT TriCam detectors as a variety of morphologies, from diffuse to constrained to three pixels, circular or elongated, the detections should be morphologically similar within a tracklet, as each tracklet generally is registered on a single CCD quadrant.  As the stationary source cleaning is not reliable for detections fainter than 20.7 mag (the Gaia limit), tracklets with a detection fainter than that threshold must be visually checked against a Sloan Digital Sky Survey thumbnail or other reference, provided by {\tt\string FindPOTATOs}  output, to ensure that none of the tracklet detections are stars. On average, $65\%$ of tracklets were judged to be legitimate. 

New tracklets are automatically processed by the MPC ingest pipelines. As part of the normal MPC processing, since these measurements are from more than 20 years ago, most new tracklets were temporarily placed in the Isolated Tracklet File, a repository of tracklets not associated with known objects, and then periodically assigned to known objects over the course of several weeks by automated scripts. 

Re-measurements needed to be manually processed by MPC staff. They were submitted separately via Jira HelpDesk ticket requests, along with a file that associated each remeasured tracklet with the previously submitted tracklet. New routines were created that allowed for the replacement of the original NEAT observations with these re-measurements. The necessity of creation of new routines for observation replacement by the MPC is an indication of the novelty of the scope of this reprocessing effort.

\section{Results}

We have submitted 2441 new tracklets and 4005 re-measurements to the MPC from ten nights of observations. The original observations span the dates of 26 November 2001 to 27 January 2002. During these nights, NEAT both followed a standard NEO-search cadence with 20 or 60 s exposures and spent significant time observing the main belt with 120 s or 150 s exposures. Accordingly, most of the measurements are of main belt objects. Due to the unique features of the data, including the undocumented features such as telescope movement, this was the volume of data processable within the scope of this effort.

The re-measurements, on average, include $80\%$ of the originally submitted NEAT tracklets, though later runs increased that fraction to $90\%$. The fraction of NEAT tracklets that were not remeasured were due to proximity of stars near the tracklet, false negatives from the machine learning model, the removal of a frame due to telescope movement, and differences in judgment on if an source's morphology resembled a minor planet or an artifact. As we require tracklets to consist of three detections, a difference of one in nine detections between the two programs results in a difference of 1 in 3 tracklets. 

The coordinate separation between remeasured tracklets and originally submitted tracklets is $0.66\pm 0.31$ arcseconds, in line with what we would expect given the use of the new Gaia DR3 reference catalog.

\begin{figure}[ht!]
\plotone{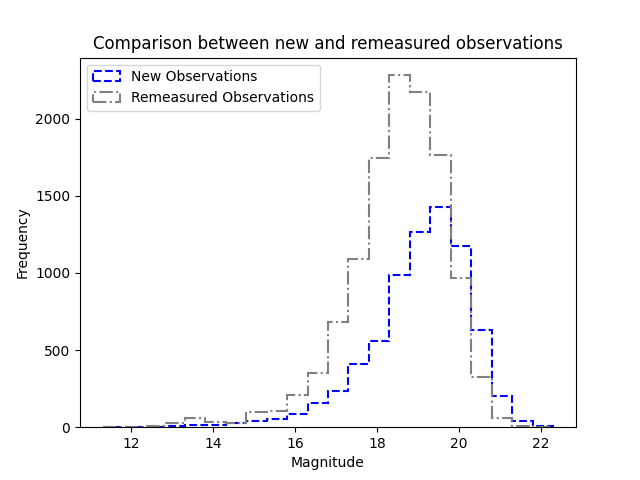}
\caption{ 
Comparison between the brightnesses of the new and remeasured observations produced by this work. The reprocessing pipeline successfully finds fainter objects, though it also finds bright (13 mag) observations not previously reported to the Minor Planet Center. 
\label{fig:mag}}
\end{figure}

\begin{figure}[ht!]
\plotone{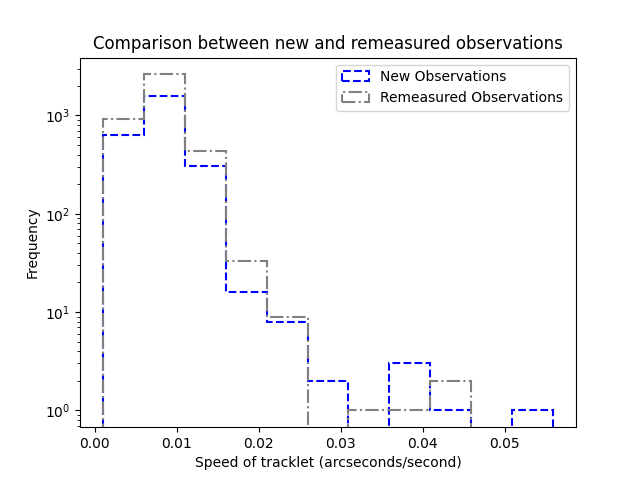}
\caption{ The speed of the new tracklets is roughly comparable in distribution to the speed of the previously submitted tracklets.
\label{fig:speed}}
\end{figure}

Compared to the re-measurements, new tracklets skew slightly fainter, though bright (13-16 mag) new tracklets were also reported (Figure \ref{fig:mag}). Although the data quality and linking software allow for the retrieval of objects moving twice as fast as the original NEAT survey, in practice, the distribution of tracklet speeds between new and remeasured tracklets is similar (Figure \ref{fig:speed}). 

Without access to the code used by the original NEAT survey, it is difficult to identify with certainty the reasons why the tracklets identified by this effort were not initially reported to the MPC. The NEAT pipeline was operated with vastly different hardware and software than is currently available and had to process images within a short window of time to enable follow-up of discoveries. Therefore, design trade-offs were made. The NEAT routine {\tt\string TABMATCH}, for example, removed objects within a 20-pixel radius of each other. This addressed the issue of false detections from star diffraction spikes \citep{1999Pravdo}. This may have eliminated some tracklets that were found by this work; indeed, new tracklets often have a detection that is within 20 pixels of a star. 

We have reported precovery observations of objects to the Minor Planet Center. For example, the previous earliest observation of (669498) 2012 XO104, a Hungaria, was from 6 December 2012. We reported observations from 02 January 2002. As of the submission of this manuscript, the most recent observation of this object is from 12 May 2024, and this object is currently unobservable. Precoveries from this work, as it continues, can be tracked via the MPEC Watch service \citep{2023Ye} by selecting the NEAT survey and years 2024 and later.\footnote{\url{https://sbnmpc.astro.umd.edu/mpecwatch/} Note observations of objects that occurred prior to the object's discovery (often termed precoveries) can be listed as recoveries on this site, since they were published after the MPEC discovery announcement.}

Some new tracklets remain in the Isolated Tracklet File, even after the completion of automated scripts by the Minor Planet Center, meaning that they could not be linked to known objects. It is possible that as more NEAT observation nights are processed, follow up observations will be obtained for those isolated tracklets.

\section{Conclusions}

This is the first documented effort to reprocess asteroid discovery survey images on a large scale, where a significant number of observations were remeasured using an updated star catalog and thousands of new tracklets were found and reported to the MPC.  For every two tracklets submitted by the original NEAT survey, we are submitting one new tracklet, in addition to remeasured tracklets.  

The new tracklets skew fainter, though new bright tracklets are also recovered.  This work is extending observed arcs, including those of currently unobservable objects. 

This pipeline will be applied to additional nights of NEAT images. The total estimated data volume of the archived NEAT data on the Small Bodies Node is 1.99 TB, comprised of $10^{5}$ images. The original number of minor planet detections submitted to the MPC from these images is 1,497,875. 

Although archival data are a powerful resource, special care must be taken to ensure accuracy. Some modern software tools are not well adapted to these data and can return misleading results. Specifically, the distortion due to the Schmidt telescope and the four readout amplifiers in the camera means that astrometric calibration must be done on each CCD quadrant independently before astrometry can be reported. Without careful calibration, users may report erroneous positions to the Minor Planet Center. 

\begin{acknowledgments}
\section{Acknowledgments}
Support for this work was provided by NASA grant 80NSSC19K1585. The authors are deeply grateful to the Massachusetts Green High Performance Computing Center (MGHPCC) operated by Boston University for providing the computational resources to complete much of this work. We are grateful to the anonymous referees, whose thoughtful and careful comments significantly improved the manuscript. This research made use of Photutils, an Astropy package for detection and photometry of astronomical sources (Bradley et al. 2023). This research has made use of the NASA/IPAC Infrared Science Archive, which is funded by the National Aeronautics and Space Administration and operated by the California Institute of Technology. This research has made use of IMCCE's SkyBoT VO tool. Data from the MPC's database is made freely available to the public. Funding for the MPC's operations comes from a NASA PDCO grant (80NSSC22M0024), administered via a University of Maryland - SAO subaward (106075-Z6415201). The MPC's computing equipment is funded in part by the above award, and in part by funding from the Tamkin Foundation. Some student research funding support also came from the Massachusetts Space Grant program as well as Olin College. This work was also supported by a Fulbright Denmark US Scholar Grant. 
\end{acknowledgments}

\software{astropy \citep{2022Astropy, 2018Astropy, 2013Astropy},  
          astroquery \citep{2019Ginsburg},
          photutils \citep{2023Bradley},
          sklearn \citep{2011Pedregosa},
          Astrometry.net \citep{2010Lang},
          SkyBoT \citep{2006Berthier},
          FindPOTATOS \citep{2025Nugent} 
          }

\bibliography{NEAT_Bib}{}

\begin{thebibliography}{}
\expandafter\ifx\csname natexlab\endcsname\relax\def\natexlab#1{#1}\fi
\providecommand{\url}[1]{\href{#1}{#1}}
\providecommand{\dodoi}[1]{doi:~\href{http://doi.org/#1}{\nolinkurl{#1}}}
\providecommand{\doeprint}[1]{\href{http://ascl.net/#1}{\nolinkurl{http://ascl.net/#1}}}
\providecommand{\doarXiv}[1]{\href{https://arxiv.org/abs/#1}{\nolinkurl{https://arxiv.org/abs/#1}}}

\bibitem[{{Astropy Collaboration} {et~al.}(2013){Astropy Collaboration},
  {Robitaille}, {Tollerud}, {Greenfield}, {Droettboom}, {Bray}, {Aldcroft},
  {Davis}, {Ginsburg}, {Price-Whelan}, {Kerzendorf}, {Conley}, {Crighton},
  {Barbary}, {Muna}, {Ferguson}, {Grollier}, {Parikh}, {Nair}, {Unther},
  {Deil}, {Woillez}, {Conseil}, {Kramer}, {Turner}, {Singer}, {Fox}, {Weaver},
  {Zabalza}, {Edwards}, {Azalee Bostroem}, {Burke}, {Casey}, {Crawford},
  {Dencheva}, {Ely}, {Jenness}, {Labrie}, {Lim}, {Pierfederici}, {Pontzen},
  {Ptak}, {Refsdal}, {Servillat}, \& {Streicher}}]{2013Astropy}
{Astropy Collaboration}, {Robitaille}, T.~P., {Tollerud}, E.~J., {et~al.} 2013,
  {Astropy: A community Python package for astronomy}, \aap, 558, A33,
  \dodoi{10.1051/0004-6361/201322068}

\bibitem[{{Astropy Collaboration} {et~al.}(2018){Astropy Collaboration},
  {Price-Whelan}, {Sip{\H{o}}cz}, {G{\"u}nther}, {Lim}, {Crawford}, {Conseil},
  {Shupe}, {Craig}, {Dencheva}, {Ginsburg}, {VanderPlas}, {Bradley},
  {P{\'e}rez-Su{\'a}rez}, {de Val-Borro}, {Aldcroft}, {Cruz}, {Robitaille},
  {Tollerud}, {Ardelean}, {Babej}, {Bach}, {Bachetti}, {Bakanov}, {Bamford},
  {Barentsen}, {Barmby}, {Baumbach}, {Berry}, {Biscani}, {Boquien}, {Bostroem},
  {Bouma}, {Brammer}, {Bray}, {Breytenbach}, {Buddelmeijer}, {Burke},
  {Calderone}, {Cano Rodr{\'\i}guez}, {Cara}, {Cardoso}, {Cheedella}, {Copin},
  {Corrales}, {Crichton}, {D'Avella}, {Deil}, {Depagne}, {Dietrich}, {Donath},
  {Droettboom}, {Earl}, {Erben}, {Fabbro}, {Ferreira}, {Finethy}, {Fox},
  {Garrison}, {Gibbons}, {Goldstein}, {Gommers}, {Greco}, {Greenfield},
  {Groener}, {Grollier}, {Hagen}, {Hirst}, {Homeier}, {Horton}, {Hosseinzadeh},
  {Hu}, {Hunkeler}, {Ivezi{\'c}}, {Jain}, {Jenness}, {Kanarek}, {Kendrew},
  {Kern}, {Kerzendorf}, {Khvalko}, {King}, {Kirkby}, {Kulkarni}, {Kumar},
  {Lee}, {Lenz}, {Littlefair}, {Ma}, {Macleod}, {Mastropietro}, {McCully},
  {Montagnac}, {Morris}, {Mueller}, {Mumford}, {Muna}, {Murphy}, {Nelson},
  {Nguyen}, {Ninan}, {N{\"o}the}, {Ogaz}, {Oh}, {Parejko}, {Parley}, {Pascual},
  {Patil}, {Patil}, {Plunkett}, {Prochaska}, {Rastogi}, {Reddy Janga},
  {Sabater}, {Sakurikar}, {Seifert}, {Sherbert}, {Sherwood-Taylor}, {Shih},
  {Sick}, {Silbiger}, {Singanamalla}, {Singer}, {Sladen}, {Sooley},
  {Sornarajah}, {Streicher}, {Teuben}, {Thomas}, {Tremblay}, {Turner},
  {Terr{\'o}n}, {van Kerkwijk}, {de la Vega}, {Watkins}, {Weaver}, {Whitmore},
  {Woillez}, {Zabalza}, \& {Astropy Contributors}}]{2018Astropy}
{Astropy Collaboration}, {Price-Whelan}, A.~M., {Sip{\H{o}}cz}, B.~M., {et~al.}
  2018, {The Astropy Project: Building an Open-science Project and Status of
  the v2.0 Core Package}, \aj, 156, 123, \dodoi{10.3847/1538-3881/aabc4f}

\bibitem[{{Astropy Collaboration} {et~al.}(2022){Astropy Collaboration},
  {Price-Whelan}, {Lim}, {Earl}, {Starkman}, {Bradley}, {Shupe}, {Patil},
  {Corrales}, {Brasseur}, {N{\"o}the}, {Donath}, {Tollerud}, {Morris},
  {Ginsburg}, {Vaher}, {Weaver}, {Tocknell}, {Jamieson}, {van Kerkwijk},
  {Robitaille}, {Merry}, {Bachetti}, {G{\"u}nther}, {Aldcroft},
  {Alvarado-Montes}, {Archibald}, {B{\'o}di}, {Bapat}, {Barentsen},
  {Baz{\'a}n}, {Biswas}, {Boquien}, {Burke}, {Cara}, {Cara}, {Conroy},
  {Conseil}, {Craig}, {Cross}, {Cruz}, {D'Eugenio}, {Dencheva}, {Devillepoix},
  {Dietrich}, {Eigenbrot}, {Erben}, {Ferreira}, {Foreman-Mackey}, {Fox},
  {Freij}, {Garg}, {Geda}, {Glattly}, {Gondhalekar}, {Gordon}, {Grant},
  {Greenfield}, {Groener}, {Guest}, {Gurovich}, {Handberg}, {Hart},
  {Hatfield-Dodds}, {Homeier}, {Hosseinzadeh}, {Jenness}, {Jones}, {Joseph},
  {Kalmbach}, {Karamehmetoglu}, {Ka{\l}uszy{\'n}ski}, {Kelley}, {Kern},
  {Kerzendorf}, {Koch}, {Kulumani}, {Lee}, {Ly}, {Ma}, {MacBride}, {Maljaars},
  {Muna}, {Murphy}, {Norman}, {O'Steen}, {Oman}, {Pacifici}, {Pascual},
  {Pascual-Granado}, {Patil}, {Perren}, {Pickering}, {Rastogi}, {Roulston},
  {Ryan}, {Rykoff}, {Sabater}, {Sakurikar}, {Salgado}, {Sanghi}, {Saunders},
  {Savchenko}, {Schwardt}, {Seifert-Eckert}, {Shih}, {Jain}, {Shukla}, {Sick},
  {Simpson}, {Singanamalla}, {Singer}, {Singhal}, {Sinha}, {Sip{\H{o}}cz},
  {Spitler}, {Stansby}, {Streicher}, {{\v{S}}umak}, {Swinbank}, {Taranu},
  {Tewary}, {Tremblay}, {de Val-Borro}, {Van Kooten}, {Vasovi{\'c}}, {Verma},
  {de Miranda Cardoso}, {Williams}, {Wilson}, {Winkel}, {Wood-Vasey}, {Xue},
  {Yoachim}, {Zhang}, {Zonca}, \& {Astropy Project Contributors}}]{2022Astropy}
{Astropy Collaboration}, {Price-Whelan}, A.~M., {Lim}, P.~L., {et~al.} 2022,
  {The Astropy Project: Sustaining and Growing a Community-oriented Open-source
  Project and the Latest Major Release (v5.0) of the Core Package}, \apj, 935,
  167, \dodoi{10.3847/1538-4357/ac7c74}

\bibitem[{{Bauer} \& {Lawrence}(2013)}]{NEATSBN}
{Bauer}, J.~M., \& {Lawrence}, K.~J. 2013, {Near Earth Asteroid Tracking Sample
  Delivery V1.0}, NASA Planetary Data System, id. EAR-A-I1063-3-NEATSAMPLE-V1.0

\bibitem[{{Berthier} {et~al.}(2006){Berthier}, {Vachier}, {Thuillot},
  {Fernique}, {Ochsenbein}, {Genova}, {Lainey}, \& {Arlot}}]{2006Berthier}
{Berthier}, J., {Vachier}, F., {Thuillot}, W., {et~al.} 2006, in Astronomical
  Society of the Pacific Conference Series, Vol. 351, Astronomical Data
  Analysis Software and Systems XV, ed. C.~{Gabriel}, C.~{Arviset}, D.~{Ponz},
  \& S.~{Enrique}, 367

\bibitem[{{Bertin} \& {Arnouts}(1996)}]{1996Bertin}
{Bertin}, E., \& {Arnouts}, S. 1996, {SExtractor: Software for source
  extraction.}, \aaps, 117, 393, \dodoi{10.1051/aas:1996164}

\bibitem[{{Boattini} {et~al.}(2001){Boattini}, {D'Abramo}, {Forti}, \&
  {Gal}}]{2001Boattini}
{Boattini}, A., {D'Abramo}, G., {Forti}, G., \& {Gal}, R. 2001, {The Arcetri
  NEO Precovery Program}, \aap, 375, 293, \dodoi{10.1051/0004-6361:20010825}

\bibitem[{Bradley {et~al.}(2023)Bradley, Sipőcz, Robitaille, Tollerud,
  Vinícius, Deil, Barbary, Wilson, Busko, Donath, Günther, Cara, Lim,
  Meßlinger, Conseil, Burnett, Bostroem, Droettboom, Bray, Bratholm, Jamieson,
  Ginsburg, Barentsen, Craig, Morris, Perrin, Rathi, Pascual, Perren, \&
  Georgiev}]{2023Bradley}
Bradley, L., Sipőcz, B., Robitaille, T., {et~al.} 2023, astropy/photutils:
  1.10.0, 1.10.0,  Zenodo, \dodoi{10.5281/zenodo.1035865}

\bibitem[{{Gaia Collaboration} {et~al.}(2023){Gaia Collaboration}, {Vallenari},
  {Brown}, {Prusti}, {de Bruijne}, {Arenou}, {Babusiaux}, {Biermann},
  {Creevey}, {Ducourant}, {Evans}, {Eyer}, {Guerra}, {Hutton}, {Jordi},
  {Klioner}, {Lammers}, {Lindegren}, {Luri}, {Mignard}, {Panem}, {Pourbaix},
  {Randich}, {Sartoretti}, {Soubiran}, {Tanga}, {Walton}, {Bailer-Jones},
  {Bastian}, {Drimmel}, {Jansen}, {Katz}, {Lattanzi}, {van Leeuwen}, {Bakker},
  {Cacciari}, {Casta{\~n}eda}, {De Angeli}, {Fabricius}, {Fouesneau},
  {Fr{\'e}mat}, {Galluccio}, {Guerrier}, {Heiter}, {Masana}, {Messineo},
  {Mowlavi}, {Nicolas}, {Nienartowicz}, {Pailler}, {Panuzzo}, {Riclet}, {Roux},
  {Seabroke}, {Sordo}, {Th{\'e}venin}, {Gracia-Abril}, {Portell}, {Teyssier},
  {Altmann}, {Andrae}, {Audard}, {Bellas-Velidis}, {Benson}, {Berthier},
  {Blomme}, {Burgess}, {Busonero}, {Busso}, {C{\'a}novas}, {Carry}, {Cellino},
  {Cheek}, {Clementini}, {Damerdji}, {Davidson}, {de Teodoro}, {Nu{\~n}ez
  Campos}, {Delchambre}, {Dell'Oro}, {Esquej}, {Fern{\'a}ndez-Hern{\'a}ndez},
  {Fraile}, {Garabato}, {Garc{\'\i}a-Lario}, {Gosset}, {Haigron}, {Halbwachs},
  {Hambly}, {Harrison}, {Hern{\'a}ndez}, {Hestroffer}, {Hodgkin}, {Holl},
  {Jan{\ss}en}, {Jevardat de Fombelle}, {Jordan}, {Krone-Martins}, {Lanzafame},
  {L{\"o}ffler}, {Marchal}, {Marrese}, {Moitinho}, {Muinonen}, {Osborne},
  {Pancino}, {Pauwels}, {Recio-Blanco}, {Reyl{\'e}}, {Riello}, {Rimoldini},
  {Roegiers}, {Rybizki}, {Sarro}, {Siopis}, {Smith}, {Sozzetti}, {Utrilla},
  {van Leeuwen}, {Abbas}, {{\'A}brah{\'a}m}, {Abreu Aramburu}, {Aerts},
  {Aguado}, {Ajaj}, {Aldea-Montero}, {Altavilla}, {{\'A}lvarez}, {Alves},
  {Anders}, {Anderson}, {Anglada Varela}, {Antoja}, {Baines}, {Baker},
  {Balaguer-N{\'u}{\~n}ez}, {Balbinot}, {Balog}, {Barache}, {Barbato},
  {Barros}, {Barstow}, {Bartolom{\'e}}, {Bassilana}, {Bauchet}, {Becciani},
  {Bellazzini}, {Berihuete}, {Bernet}, {Bertone}, {Bianchi}, {Binnenfeld},
  {Blanco-Cuaresma}, {Blazere}, {Boch}, {Bombrun}, {Bossini}, {Bouquillon},
  {Bragaglia}, {Bramante}, {Breedt}, {Bressan}, {Brouillet}, {Brugaletta},
  {Bucciarelli}, {Burlacu}, {Butkevich}, {Buzzi}, {Caffau}, {Cancelliere},
  {Cantat-Gaudin}, {Carballo}, {Carlucci}, {Carnerero}, {Carrasco},
  {Casamiquela}, {Castellani}, {Castro-Ginard}, {Chaoul}, {Charlot}, {Chemin},
  {Chiaramida}, {Chiavassa}, {Chornay}, {Comoretto}, {Contursi}, {Cooper},
  {Cornez}, {Cowell}, {Crifo}, {Cropper}, {Crosta}, {Crowley}, {Dafonte},
  {Dapergolas}, {David}, {David}, {de Laverny}, {De Luise}, {De March}, {De
  Ridder}, {de Souza}, {de Torres}, {del Peloso}, {del Pozo}, {Delbo},
  {Delgado}, {Delisle}, {Demouchy}, {Dharmawardena}, {Di Matteo}, {Diakite},
  {Diener}, {Distefano}, {Dolding}, {Edvardsson}, {Enke}, {Fabre}, {Fabrizio},
  {Faigler}, {Fedorets}, {Fernique}, {Fienga}, {Figueras}, {Fournier},
  {Fouron}, {Fragkoudi}, {Gai}, {Garcia-Gutierrez}, {Garcia-Reinaldos},
  {Garc{\'\i}a-Torres}, {Garofalo}, {Gavel}, {Gavras}, {Gerlach}, {Geyer},
  {Giacobbe}, {Gilmore}, {Girona}, {Giuffrida}, {Gomel}, {Gomez},
  {Gonz{\'a}lez-N{\'u}{\~n}ez}, {Gonz{\'a}lez-Santamar{\'\i}a},
  {Gonz{\'a}lez-Vidal}, {Granvik}, {Guillout}, {Guiraud},
  {Guti{\'e}rrez-S{\'a}nchez}, {Guy}, {Hatzidimitriou}, {Hauser}, {Haywood},
  {Helmer}, {Helmi}, {Sarmiento}, {Hidalgo}, {Hilger}, {H{\l}adczuk}, {Hobbs},
  {Holland}, {Huckle}, {Jardine}, {Jasniewicz}, {Jean-Antoine Piccolo},
  {Jim{\'e}nez-Arranz}, {Jorissen}, {Juaristi Campillo}, {Julbe}, {Karbevska},
  {Kervella}, {Khanna}, {Kontizas}, {Kordopatis}, {Korn}, {K{\'o}sp{\'a}l},
  {Kostrzewa-Rutkowska}, {Kruszy{\'n}ska}, {Kun}, {Laizeau}, {Lambert},
  {Lanza}, {Lasne}, {Le Campion}, {Lebreton}, {Lebzelter}, {Leccia}, {Leclerc},
  {Lecoeur-Taibi}, {Liao}, {Licata}, {Lindstr{\o}m}, {Lister}, {Livanou},
  {Lobel}, {Lorca}, {Loup}, {Madrero Pardo}, {Magdaleno Romeo}, {Managau},
  {Mann}, {Manteiga}, {Marchant}, {Marconi}, {Marcos}, {Marcos Santos},
  {Mar{\'\i}n Pina}, {Marinoni}, {Marocco}, {Marshall}, {Martin Polo},
  {Mart{\'\i}n-Fleitas}, {Marton}, {Mary}, {Masip}, {Massari},
  {Mastrobuono-Battisti}, {Mazeh}, {McMillan}, {Messina}, {Michalik}, {Millar},
  {Mints}, {Molina}, {Molinaro}, {Moln{\'a}r}, {Monari}, {Mongui{\'o}},
  {Montegriffo}, {Montero}, {Mor}, {Mora}, {Morbidelli}, {Morel}, {Morris},
  {Muraveva}, {Murphy}, {Musella}, {Nagy}, {Noval}, {Oca{\~n}a}, {Ogden},
  {Ordenovic}, {Osinde}, {Pagani}, {Pagano}, {Palaversa}, {Palicio},
  {Pallas-Quintela}, {Panahi}, {Payne-Wardenaar}, {Pe{\~n}alosa Esteller},
  {Penttil{\"a}}, {Pichon}, {Piersimoni}, {Pineau}, {Plachy}, {Plum}, {Poggio},
  {Pr{\v{s}}a}, {Pulone}, {Racero}, {Ragaini}, {Rainer}, {Raiteri}, {Rambaux},
  {Ramos}, {Ramos-Lerate}, {Re Fiorentin}, {Regibo}, {Richards}, {Rios Diaz},
  {Ripepi}, {Riva}, {Rix}, {Rixon}, {Robichon}, {Robin}, {Robin}, {Roelens},
  {Rogues}, {Rohrbasser}, {Romero-G{\'o}mez}, {Rowell}, {Royer}, {Ruz Mieres},
  {Rybicki}, {Sadowski}, {S{\'a}ez N{\'u}{\~n}ez}, {Sagrist{\`a} Sell{\'e}s},
  {Sahlmann}, {Salguero}, {Samaras}, {Sanchez Gimenez}, {Sanna},
  {Santove{\~n}a}, {Sarasso}, {Schultheis}, {Sciacca}, {Segol}, {Segovia},
  {S{\'e}gransan}, {Semeux}, {Shahaf}, {Siddiqui}, {Siebert}, {Siltala},
  {Silvelo}, {Slezak}, {Slezak}, {Smart}, {Snaith}, {Solano}, {Solitro},
  {Souami}, {Souchay}, {Spagna}, {Spina}, {Spoto}, {Steele},
  {Steidelm{\"u}ller}, {Stephenson}, {S{\"u}veges}, {Surdej}, {Szabados},
  {Szegedi-Elek}, {Taris}, {Taylor}, {Teixeira}, {Tolomei}, {Tonello}, {Torra},
  {Torra}, {Torralba Elipe}, {Trabucchi}, {Tsounis}, {Turon}, {Ulla}, {Unger},
  {Vaillant}, {van Dillen}, {van Reeven}, {Vanel}, {Vecchiato}, {Viala},
  {Vicente}, {Voutsinas}, {Weiler}, {Wevers}, {Wyrzykowski}, {Yoldas}, {Yvard},
  {Zhao}, {Zorec}, {Zucker}, \& {Zwitter}}]{2023Gaia}
{Gaia Collaboration}, {Vallenari}, A., {Brown}, A.~G.~A., {et~al.} 2023, {Gaia
  Data Release 3. Summary of the content and survey properties}, \aap, 674, A1,
  \dodoi{10.1051/0004-6361/202243940}

\bibitem[{{Gehrels} {et~al.}(1986){Gehrels}, {Marsden}, {McMillan}, \&
  {Scotti}}]{1986Gehrels}
{Gehrels}, T., {Marsden}, B.~G., {McMillan}, R.~S., \& {Scotti}, J.~V. 1986,
  {Astrometry with a scanning CCD}, \aj, 91, 1242, \dodoi{10.1086/114098}

\bibitem[{{Gehrels} \& {McMillan}(1982)}]{1982Gehrels}
{Gehrels}, T., \& {McMillan}, R.~S. 1982, in Astrophysics and Space Science
  Library, Vol.~96, Sun and Planetary System, ed. W.~{Fricke} \& G.~{Teleki},
  279, \dodoi{10.1007/978-94-009-7846-1_75}

\bibitem[{{Ginsburg} {et~al.}(2019){Ginsburg}, {Sip{\H{o}}cz}, {Brasseur},
  {Cowperthwaite}, {Craig}, {Deil}, {Guillochon}, {Guzman}, {Liedtke}, {Lian
  Lim}, {Lockhart}, {Mommert}, {Morris}, {Norman}, {Parikh}, {Persson},
  {Robitaille}, {Segovia}, {Singer}, {Tollerud}, {de Val-Borro}, {Valtchanov},
  {Woillez}, {Astroquery Collaboration}, \& {a subset of astropy
  Collaboration}}]{2019Ginsburg}
{Ginsburg}, A., {Sip{\H{o}}cz}, B.~M., {Brasseur}, C.~E., {et~al.} 2019,
  {astroquery: An Astronomical Web-querying Package in Python}, \aj, 157, 98,
  \dodoi{10.3847/1538-3881/aafc33}

\bibitem[{{Granvik} {et~al.}(2018){Granvik}, {Morbidelli}, {Jedicke}, {Bolin},
  {Bottke}, {Beshore}, {Vokrouhlick{\'y}}, {Nesvorn{\'y}}, \&
  {Michel}}]{2018Granvik}
{Granvik}, M., {Morbidelli}, A., {Jedicke}, R., {et~al.} 2018, {Debiased orbit
  and absolute-magnitude distributions for near-Earth objects}, \icarus, 312,
  181, \dodoi{10.1016/j.icarus.2018.04.018}

\bibitem[{{Gwyn} {et~al.}(2012){Gwyn}, {Hill}, \& {Kavelaars}}]{2012Gwyn}
{Gwyn}, S. D.~J., {Hill}, N., \& {Kavelaars}, J.~J. 2012, {SSOS: A
  Moving-Object Image Search Tool for Asteroid Precovery}, \pasp, 124, 579,
  \dodoi{10.1086/666462}

\bibitem[{{Helin} \& {Dunbar}(1990)}]{1990Helin}
{Helin}, E.~F., \& {Dunbar}, R.~S. 1990, {Search Techniques for Near Earth
  Asteroids}, Vistas in Astronomy, 33, 21, \dodoi{10.1016/0083-6656(90)90030-C}

\bibitem[{{Helin} {et~al.}(1997){Helin}, {Pravdo}, {Rabinowitz}, \&
  {Lawrence}}]{1997Helin}
{Helin}, E.~F., {Pravdo}, S.~H., {Rabinowitz}, D.~L., \& {Lawrence}, K.~J.
  1997, {Near-Earth Asteroid Tracking (NEAT) Program}, Annals of the New York
  Academy of Sciences, 822, 6, \dodoi{10.1111/j.1749-6632.1997.tb48329.x}

\bibitem[{{Jedicke} {et~al.}(2015){Jedicke}, {Granvik}, {Micheli}, {Ryan},
  {Spahr}, \& {Yeomans}}]{2015Jedicke}
{Jedicke}, R., {Granvik}, M., {Micheli}, M., {et~al.} 2015, in Asteroids IV,
  ed. P.~{Michel}, F.~E. {DeMeo}, \& W.~F. {Bottke} (University of Arizona
  Press), 795--813, \dodoi{10.2458/azu_uapress_9780816532131-ch040}

\bibitem[{{Lang} {et~al.}(2010){Lang}, {Hogg}, {Mierle}, {Blanton}, \&
  {Roweis}}]{2010Lang}
{Lang}, D., {Hogg}, D.~W., {Mierle}, K., {Blanton}, M., \& {Roweis}, S. 2010,
  {Astrometry.net: Blind Astrometric Calibration of Arbitrary Astronomical
  Images}, \aj, 139, 1782, \dodoi{10.1088/0004-6256/139/5/1782}

\bibitem[{{M}c{K}inney(2010)}]{2010mckinney}
{M}c{K}inney, W. 2010, in {P}roceedings of the 9th {P}ython in {S}cience
  {C}onference, ed. {S}t\'efan van~der {W}alt \& {J}arrod {M}illman, 56 -- 61,
  \dodoi{10.25080/Majora-92bf1922-00a}

\bibitem[{{Nesvorn{\'y}} {et~al.}(2024){Nesvorn{\'y}}, {Vokrouhlick{\'y}},
  {Shelly}, {Deienno}, {Bottke}, {Fuls}, {Jedicke}, {Naidu}, {Chesley},
  {Chodas}, {Farnocchia}, \& {Delbo}}]{2024Nesvorny}
{Nesvorn{\'y}}, D., {Vokrouhlick{\'y}}, D., {Shelly}, F., {et~al.} 2024,
  {NEOMOD 3: The debiased size distribution of Near Earth Objects}, \icarus,
  417, 116110, \dodoi{10.1016/j.icarus.2024.116110}

\bibitem[{{Nugent} {et~al.}(2025){Nugent}, {Tan}, \& {Bauer}}]{2025Nugent}
{Nugent}, C.~R., {Tan}, N., \& {Bauer}, J. 2025, {FINDPOTATOs: Minor Planet
  Observation Linking Software}, The Planetary Science Journal,
  \dodoi{10.3847/PSJ/ad94eb}

\bibitem[{pandas~development team(2024)}]{pandas}
pandas~development team, T. 2024, pandas-dev/pandas: Pandas, v2.2.0,  Zenodo,
  \dodoi{10.5281/zenodo.10537285}

\bibitem[{Pedregosa {et~al.}(2011)Pedregosa, Varoquaux, Gramfort, Michel,
  Thirion, Grisel, Blondel, Prettenhofer, Weiss, Dubourg, Vanderplas, Passos,
  Cournapeau, Brucher, Perrot, \& {{\'E}}douard Duchesnay}]{2011Pedregosa}
Pedregosa, F., Varoquaux, G., Gramfort, A., {et~al.} 2011, Scikit-learn:
  Machine Learning in Python, Journal of Machine Learning Research, 12, 2825.
\newblock \url{http://jmlr.org/papers/v12/pedregosa11a.html}

\bibitem[{{Perlbarg} {et~al.}(2023){Perlbarg}, {Desmars}, {Robert}, \&
  {Hestroffer}}]{2023Perlbarg}
{Perlbarg}, A.~C., {Desmars}, J., {Robert}, V., \& {Hestroffer}, D. 2023,
  {NAROO program. Precovery observations of potentially hazardous asteroids},
  \aap, 680, A41, \dodoi{10.1051/0004-6361/202347100}

\bibitem[{{Pravdo} {et~al.}(1999){Pravdo}, {Rabinowitz}, {Helin}, {Lawrence},
  {Bambery}, {Clark}, {Groom}, {Levin}, {Lorre}, {Shaklan}, {Kervin},
  {Africano}, {Sydney}, \& {Soohoo}}]{1999Pravdo}
{Pravdo}, S.~H., {Rabinowitz}, D.~L., {Helin}, E.~F., {et~al.} 1999, {The
  Near-Earth Asteroid Tracking (NEAT) Program: an Automated System for
  Telescope Control, Wide-Field Imaging, and Object Detection}, \aj, 117, 1616,
  \dodoi{10.1086/300769}

\bibitem[{{Rabinowitz}(1991)}]{1991Rabinowitz}
{Rabinowitz}, D.~L. 1991, {Detection of earth-approaching asteroids in near
  real time}, \aj, 101, 1518, \dodoi{10.1086/115785}

\bibitem[{{Saifollahi} {et~al.}(2023){Saifollahi}, {Verdoes Kleijn},
  {Williams}, {Micheli}, {Santana-Ros}, {Helmich}, {Koschny}, \&
  {Conversi}}]{2023Saifollahi}
{Saifollahi}, T., {Verdoes Kleijn}, G., {Williams}, R., {et~al.} 2023, {Mining
  archival data from wide-field astronomical surveys in search of near-Earth
  objects}, \aap, 673, A93, \dodoi{10.1051/0004-6361/202345957}

\bibitem[{{Stetson}(1987)}]{1987Stetson}
{Stetson}, P.~B. 1987, {DAOPHOT: A Computer Program for Crowded-Field Stellar
  Photometry}, \pasp, 99, 191, \dodoi{10.1086/131977}

\bibitem[{{Vaduvescu} {et~al.}(2013){Vaduvescu}, {Popescu}, {Comsa},
  {Paraschiv}, {Lacatus}, {Sonka}, {Tudorica}, {Birlan}, {Suciu}, {Char},
  {Constantinescu}, {Badescu}, {Badea}, {Vidican}, \&
  {Opriseanu}}]{2013Vaduvescu}
{Vaduvescu}, O., {Popescu}, M., {Comsa}, I., {et~al.} 2013, {Mining the ESO WFI
  and INT WFC archives for known Near Earth Asteroids. Mega-Precovery
  software}, Astronomische Nachrichten, 334, 718,
  \dodoi{10.1002/asna.201211720}

\bibitem[{{Ye} {et~al.}(2023){Ye}, {Hibbitts}, \& {Bauer}}]{2023Ye}
{Ye}, Q.~Z., {Hibbitts}, T., \& {Bauer}, J. 2023, in LPI Contributions, Vol.
  2991, 6th Planetary Data Workshop, 7021

\end{thebibliography}
\bibliographystyle{psj}

\end{document}